
\magnification=1200
\hsize 5 in
\vsize 7 in
\nopagenumbers
\def\wt{\widetilde}
\def\wh{\widehat}

\line{Preprint {\bf SB/F/95-227}}
\line{\hfil {\bf UCVFC-DF/4-95}}
\hrule
\vskip 2cm
\centerline{\bf NON-ABELIAN ``SELF-DUAL'' MASSIVE GAUGE THEORY}
\centerline{\bf IN $2+1$ DIMENSIONS}
\vskip 1cm
\centerline{\bf P.J.Arias${}^a$,L.Leal${}^a$,A.Restuccia${}^b$}
\centerline{\it ${}^a$ Departamento de F\'{\i}sica, Facultad de Ciencias}
\centerline{\it Universidad Central de Venezuela, AP 47586, Caracas 1041-A}
\centerline{\it Venezuela}
\centerline{\it ${}^b$ Universidad Sim\'on Bol\'{\i}var, Departamento de
F\'{\i}sica}
\centerline{\it AP 89000, Caracas 1080-A}
\centerline{\it Venezuela}
\centerline{\tt e-mail: parias@tierra.ciens.ucv.ve, lleal@dino.conicit.ve}
\centerline{\tt arestu@usb.ve}
\vskip 2cm

\noindent
{\bf ABSTRACT}

\noindent {\narrower
A non-abelian ``self-dual'' massive gauge theory describing a massive spin one
physical mode is presented. The action is expressed in terms of two independent
connections on a principle bundle over $2+1$ space-time. The kinetic terms are
respectively a Chern-Simons and  a $BF$ lagrangians, while the gauge invariant
interaction is expressed in terms of the difference of the two connections. At
the linearized level the theory is equivalent to the Topological Massive gauge
theory. The covariant, $BRST$ invariant, effective action of the non-abelian
sefl dual gauge theory is constructed. \par}
\vskip 2cm
\hrule
\vfill \eject

Topological gauge invariant theories describe field theories without local
propagating physical modes. Thus, physical observables are only topological
geometrical objects. However, they have a very rich global structure allowing,
in particular, to obtain the Donaldson invariants of four manifolds [1] and
more
recently the Seiberg-Witten invariants [2]. These topological gauge invariant
terms,
when interacting with physical action lagrangians, or even between themselves
through non-
topological interaction terms, provide very special properties to the new
theories.
This effect is particulary interesting in three dimensional space-time where
the Chern-Simons ($CS$) topological terms provide masses to the gauge fields
 of vector and tensor gauge theories [3]. In the same context, point particles,
 in $2+1$ dimensions, minimally coupled to a gauge field, whose dynamics is
given
  by the $CS$ term, can acquire fractional spin and statistics [4]. This
picture
  can be generalized to $3+1$ dimensions, where the topological term is now a
$BF$
   gauge invariant one, giving rise to the so called ``string'' fractional
statistics [5].

We are going to propose here a non-abelian massive gauge theory in $2+1$
dimensions whose field equations are described by a Lorentz covariant first
order dynamical system in distinction to the second order field equation  of
the
Topological Massive ($TM$) gauge theory [3]. It is the non-abelian
generalization of
the so called ``self-dual'' ($SD$) abelian theory [6,7]. The $TM$ non-abelian
gauge theory was
formulated in [3] , the interest of the non-abelian generalization of the $SD$
theory was raised in [7] . The action of the theory we propose can be
interpreted as the interaction of a $CS$ and a $BF$  topological action terms .
However it describes a locally propagating  spin 1 physical mode as the $TM$
gauge theory.

It is known that the abelian spin $1$ theory in $2+1$ dimensions may be
described by two covariant actions: the second order $TM$
 gauge action [3] and the $SD$ action [6,7]. The first order field
 equations of the $SD$ theory are a minimal realization of the Pauli-Liubansky
  and mass shell conditions for the spin $1$ representations of the Poincare
group
   in $2+1$ dimensions [8], and both theories, the $SD$ and $TM$, describe the
same
    propagating physical degrees of freedom and have identical local
properties.
    Over simply connected regions of space-time the $SD$ theory correspond to a
    gauge fixed version of the $TM$ gauge theory [7,9]. However, once we
consider a
    topological non-trivial base manifold, where the $CS$ term may globally
contribute
    to the observables of the theory, the relation between both theories is not
    of a trivial equivalence. The global difference between both theories was
shown in [10] by comparing their partitions functions. They differ by a
topological factor equal
    to the partition function of the pure $CS$ action, which may be expressed
    in terms of the topological Ray-Singer torsion.

It was also shown in [10] that the $SD$ abelian action can be slightly
modified in order to have a theory locally and globally equivalent to the $TM$
abelian action.
This was achieved by considering the action
$$
\wh {S} = \int d^3x\Bigl[{\mu^2 \over
2}(a_{\mu}-\omega_{\mu})(a^{\mu}-\omega^{\mu})-
{\mu \over 2}a_{\mu}\varepsilon^{\mu\nu\lambda}\partial_{\nu}a_{\lambda}
\Bigl], \eqno (1)
$$
where $\omega_{\mu}$ is an independent closed one form. We may remove this
restriction
on $\omega_{\mu}$ in two ways one: by introducing a $BF$ term into the action
$$
S_I  = \int d^3 x \Bigl[{\mu^2 \over
2}(a_{\mu}-\omega_{\mu})(a^{\mu}-\omega^{\mu})-
{\mu \over 2}a_{\mu}\varepsilon^{\mu\nu\lambda}\partial_{\nu}a_{\lambda}
-{\mu \over 2}\lambda_{\mu}
\varepsilon^{\mu\nu\lambda}\partial_{\nu}\omega_{\lambda}\Bigr], \eqno (2)
$$
where $\lambda_{\mu}$ is a lagrange multiplier. This action thus
describes the interaction of two topological gauge invariant terms, a $CS$
and a $BF$ one, through a gauge invariant non-topological quadratic term.
The other way to remove the restriction on $\omega_{\mu}$ seems more
economical, it results by considering the action
$$
S_{II}  = \int d^3 x \Bigl[{\mu^2 \over
2}(a_{\mu}-\omega_{\mu})(a^{\mu}-\omega^{\mu})-
{\mu \over 2}(a_{\mu}-2\omega_{\mu})\varepsilon^{\mu\nu\lambda}\partial_{\nu}
a_{\lambda}\Bigr], \eqno (3)
$$
which emerges from the master action introduced in [7]. Both actions,
$S_I$ and $S_{II}$, are locally equivalent to the $TM$ model
and the restriction on $\omega_{\mu}$ emerges as an equation of motion.

If we make the substitution $a_{\mu}-\omega_{\mu}=b_{\mu}$ in $S_{II}$ it
decouples in two terms: the $SD$ action plus the $CS$ one. So its partition
function
is: $Z_{S_{II}}=Z_{SD} Z_{CS}=Z_{TM}$ [10]. We may also ``complete squares''
in it and the $TM$ action emerges. Repeating this last procedure on
$S_I$, the $TM$ action also emerges, plus an additional decoupled $CS$ term,
so $Z_{S_I}=Z_{TM} Z_{CS}$. These results may also be proven by performing the
canonical analysis, we omit the details here.

In order to search for an non-abelian model we can take these two abelian
actions as different departure points. For
$S_{II}$ the resulting non-abelian first order action is more likely a master
action as in [7] with a second order evolving system. This is not the case for
$S_I$.
We propose, then, the following non-abelian first order gauge action for
 the spin 1 theory
$$
S_{NA} = \int d^3 x {\mu \over {g^2}}tr
\Bigl[-\mu(a_{\mu}-\omega_{\mu})(a^{\mu}-\omega^{\mu})
-\varepsilon^{\mu\nu\lambda}\bigl ( \partial_{\mu}a_{\nu}a_{\lambda}
+ {2 \over 3}a_{\mu}a_{\nu}a_{\lambda} \bigr ) -
{{\lambda_{\mu}} \over 2}
\varepsilon^{\mu \nu \lambda}G_{\nu \lambda}\Bigr], \eqno (4)
$$
where $G_{\mu \nu}=\partial_{\mu}\omega_{\nu}-\partial_{\nu}\omega_{\mu}+
[\omega_{\mu},\omega_{\nu}]$ is the Lie algebra valued $2$-form associated to
the connection $1$-form $\omega$. $a$ and $\omega$ are independent connections
 on the same principle bundle. We use the notation $\xi = g
\xi^{a}_{\mu}dx^{\mu}
 T_a$ for Lie algebra valued $1$-forms in local coordinates. $T_a$
 are antihermitian generators, $tr\bigl(T^{a}T^{b}\bigr)
 = -{1 \over 2}\delta^{ab}$; and $f^{a}_{bc}$ will denote the
 structure constants of the Lie algebra. The coupling constant
 is denoted by $g$: so ${{\mu} \over {g^2}}$ is dimensionless.
 ${\cal D}_{\mu} = \partial_{\mu} + [a_{\mu},\quad]$
 and $\wt {\cal D} = \partial_{\mu} + [\omega_{\mu},\quad]$
 are the covariant derivatives with respect to $a$ and $\omega$
  in local coordinates. The metric signature is $(+--)$.

  The action (4) is invariant up to a closed form, under the finite
  gauge transformations
  $$
  \eqalignno{
  {a \to {t^{-1}}} &at+t^{-1}dt \cr
  {\omega \to {t^{-1}}} &{\omega}t+t^{-1}dt & (5) \cr
  {\lambda \to {t^{-1}}} &{\lambda}t \cr
  }$$
  and under the infinitesimal gauge transformations
  $$
  \eqalignno{
  {a \to a} & \cr
  {\omega \to \omega} & & (6) \cr
  {\lambda \to \lambda} &-\wt {\cal D}\chi \cr
  }$$
  where $t$ is an element of the group acting on the fiber. In order to have
  exact invariance of the exponential of the action in the functional integral,
  the same quantization condition as in the $TM$ arises. That is
  $4\pi{\mu \over {g^2}}=$integer.

  The field equations obtained from (4) are
  $$
  \eqalignno{
  -\mu(a^{\lambda}-\omega^{\lambda})+\varepsilon^{\mu \nu \lambda}F_{\mu \nu}
   &=0, & (7,a) \cr
   \mu(a^{\lambda}-\omega^{\lambda})+{1 \over 2}\varepsilon^{\mu \nu \lambda}
   \wt{\cal D}_{\mu}\lambda_{\nu} &=0, & (7,b) \cr
   G_{\nu \lambda} &=0. & (7,c) \cr
   }$$
   The integrability conditions for (7,a) and (7,b)  are
   $$
   \eqalignno{
   {\cal D}_{\lambda}(a^{\lambda}-\omega^{\lambda}) &=0, & (8,a) \cr
  \wt{\cal D}_{\lambda}(a^{\lambda}-\omega^{\lambda})+{1 \over 4}
  \varepsilon^{\mu \nu \lambda}[G_{\mu \nu},\lambda_{\lambda}] &=
  \wt{\cal D}_{\lambda}(a^{\lambda}-\omega^{\lambda})=0. & (8,b) \cr
  }$$
  which represent the same condition in spite of the fact that ${\cal D}$ and
$\wt{\cal D}$ appears in (8,a) and (8,b) respectively. There is, thus, only one
restriction in (8).

  From (7,c), $\omega$ is a flat connection constrained by (8). If
  $\Omega_0$ is a flat connection, then $\wh {\Omega}_0=s^{-1} \Omega_0 s+
  s^{-1}ds$ is also flat. Given $\Omega_0$ and $a$ there is always a
  $\wh {\Omega}_0$, also flat, satisfying (8). This means that the
  space of solutions of (7,c) and (8), the flat connections satisfying
   the Lorentz type condition (8,b), is non-empty. The integrability conditions
are then
   completely satisfied without restricting the space of solutions of (7,a)
and (7,b). Equation (7,a) determines $a$, and it implies the following
   equation for the associated curvature
   $$
   {\cal D}_{\mu}f_{\nu}-{\cal D}_{\nu}f_{\mu}-\mu F_{\mu \nu}=
   {1 \over {\mu}}[f_{\mu},f_{\nu}], \eqno (9)
   $$
   where $f^{\mu}=\varepsilon^{\mu \nu \lambda}F_{\nu \lambda}$.

   Equation (7,a) is the non-abelian generalization of the $SD$ equation
([10]).
   It is equivalent, in the abelian case, with the inclusion of the closed
   $1$-form $\omega$, to the $TM$ equations. In the non-abelian case (7,a)
   implies (9) which differs from  the $TM$ equations by the term
   $[f_{\mu},f_{\nu}]$.

   It can be shown that (7,b) does not describe any propagating physical
   mode. In fact, for a given $\omega$ and $a$, (7,b) is a linear equation
   in $\lambda$ . Their solutions are obtained from a particular one by
   adding to it all the solutions of the homogeneous equation. The latter may
   be written as
   $$
   \wt d \lambda = 0, \eqno (10)
   $$
   where $\wt d$ is a exterior derivative since
   $\wt d \wt d=\bigl[\wt{\cal D}_{\mu}\wt{\cal D}_{\nu}-
   \wt{\cal D}_{\mu}\wt{\cal D}_{\nu}\bigr]dx^{\mu} \land dx^{\nu} =
   G_{\mu \nu}dx^{\mu} \land dx^{\nu}=0$.
   The most general solution to (10) is locally
   $$
   \lambda=\wt d p, \eqno (11)
   $$
   which can be absorbed by the gauge transformation (6). Consequently
$\lambda$ does not describe locally any independent propagating physical mode.

   Applying another covariant derivative on (9), and using the Bianchi
identities we obtain
   $$
   ({\cal D}^{\mu}{\cal D}_{\mu} + \mu^2)f_{\nu}=
   {1 \over 2}\varepsilon_{\nu \alpha \beta}[f^{\alpha},f^{\beta}]+
   {1 \over {\mu}}{\cal D}^{\alpha}[f_{\alpha},f_{\nu}], \eqno (12)
   $$
showing the massive character of the excitations.

The action (3) describes thus a spin 1 massive physical mode with a dynamics
given by a covariant first order evolving system.

 We are now going to construct the covariant $BRST$ invariant effective action
of the theory.

{}From (3) we see that $a_0$, $\omega_0$ and $\lambda_0$ are non-dynamical
variables . Making variations with respect to $a_0$ we will be able to
eliminate it, arriving, after the $2+1$ canonical decomposition, to
$$
\eqalignno{
S_{NA}=\int d^3x{\mu \over {g^2}}tr\Bigl[& \varepsilon_{ij}\dot {a_i}a_j+
\varepsilon_{ij} \dot {\omega_i}\lambda_j+ \cr
&+{{\lambda_0} \over 2}\varepsilon_{ij}G_{ij}+\omega_0 \varepsilon_{ij}
(F_{ij}+\wt {\cal D}_i \lambda_j)+ \cr
&+\mu(a_i -\omega_i)(a_i -\omega_i)+{1 \over {4\mu}}
\varepsilon_{ij}F_{ij}\varepsilon_{kl}F_{kl}\Bigr], & (13) \cr
}$$
where the role played by each variable is clear: $\omega_i$, $a_i$ are
the dynamical variables; $\lambda_0$ and $\omega_0$ are Lagrange multipliers
associated to the constraints
$$
\eqalignno{
\varphi &={{\mu} \over 4}\varepsilon_{ij}G_{ij}, & (14,a) \cr
\psi &={{\mu} \over 2}\varepsilon_{ij}(F_{ij}-\wt {\cal D}_i \lambda_j). &
 (14,b) \cr
 }$$
 Also from (13) we read the conjugated momenta to the dynamical variables
 $$
 \eqalignno{
 \Pi_i & \equiv gT^a{{\partial{\cal L}} \over {\partial \dot {a}_i^a}}
 =-{{\mu} \over 2}\varepsilon_{ij}a_j, & (15,a) \cr
 P_i & \equiv gT^a{{\partial{\cal L}} \over {\partial \dot {\omega}_i^a}}=
 -{{\mu} \over 2}\varepsilon_{ij}\lambda_j. & (15,b) \cr
 }$$
We will take (15,b) as a definition of $\lambda_i$ and keep the primary
constraint
$$
\psi_i \equiv \Pi_i+{\mu \over 2}\varepsilon_{ij}a_j, \eqno (16)
$$
in order to keep explicit rotational invariance. Taking (14)-(16) into account
we arrive to the total hamiltonian density [11,12]
$$
{\cal H}_T = {\cal H}_0 +\lambda_0^a \varphi^a +\omega_0^a \psi^a +
\rho_i^a \psi_i^a, \eqno (17)
$$
where
$$
{\cal H}_0={{\mu^2} \over 2}(a_i^a - \omega_i^a)(a_i^a - \omega_i^a)+
{1 \over 8}\varepsilon_{ij}F_{ij}^a\varepsilon_{kl}F_{kl}^a. \eqno (18)
$$
Following the conservation conditions, we find that there are no more
constraints, and that the first class constraints are: $\varphi^a$ (as in
(14,a)) and
$$
\eqalignno{
\theta^a & \equiv \varphi^a - {\cal D}_i \psi_i^a, \cr
& ={{\mu} \over 2}\varepsilon_{ij}F_{ij}-
{\mu \over 2}\varepsilon_{ij}{\cal D}_i a_j -
{\cal D}_i\Pi_i^a -\wt {\cal D}_i P_i. & (19) \cr
}$$
$\psi_i^a$ are second class constraints, so we must
proceed using Dirac brackets:
$$
{\bigl\{F,G\bigr\}^{\ast}\equiv
\Bigl[\bigl\{F,G\bigr\}-\int d^2x d^2y \bigl\{F,\psi_i^a(x)\bigr\}
M_{ij}^{ab}(x,y)\bigl\{\psi_j^b(y),G\bigr\}\Bigr]
{\vrule height 20 pt}}_{\psi_i^a=0}
$$
where
$$
M_{ij}^{ab}(x,y) \equiv {\bigl\{ \psi_i^a(x),\psi_j^b(y)\bigr\}}^{-1}=
-{1 \over {\mu}}\varepsilon_{ij}\delta^{ab}\delta^2(x-y). \eqno (20)
$$

The Dirac algebra between the constraints and $H_0$ is closed and has the
only non-null equal time brackets
$$
\eqalignno{
\bigl\{\varphi^a(x),\theta^b(y)\bigr\}^{\ast} &=-gf^{abc}\varphi^c(x)
\delta^2(x-y), & (21,a) \cr
\bigl\{\theta^a(x),\theta^b(y)\bigr\}^{\ast} &=-gf^{abc}\theta^c(x)
\delta^2(x-y). & (21,b) \cr
}$$

Now the $BRST$ method proceeds: the constraint algebra shows that this theory
 is of rank one and the method we will apply is equivalent to the $BFV$ one
 [13,14]. We introduce the ghosts $C^a$ and $D^a$ associated respectively
 to the first class constraints $\varphi^a$ and $\theta^a$, and a pair of
 antighost $\overline C^a$ and $\overline D^a$. Their parity is opposed
 to that of the constraints. The $BRST$ transformation is defined as
 $\delta_{BRST}F \equiv \xi \hat{\delta}F$, with
 $$
 \hat{\delta}F \equiv C^a \bigl\{F,\varphi^a\bigr\}^{\ast}+
 D^a\bigl\{F,\theta^a\bigr\}^{\ast}. \eqno (22)
 $$
 Requiring that ${\hat{\delta}}^2F=0$ we find that
 $$
 \eqalignno{
 \hat{\delta}C^a &= -gf^{abc}C^bD^c, & (23,a) \cr
 \hat{\delta}D^a &= -{g \over 2}f^{abc}D^bD^c, & (23,b) \cr
 }$$
 The effective action is then given by [14]
 $$
 \eqalignno{
 S_{eff}= \int d^3x\biggl[\Pi_i^a\dot a_i^a +P_i^a\dot{\omega}_i^a
 -{\cal H}_0 -& \lambda_0^a\varphi^a -\omega_0^a\theta^a- \cr
 &-\hat{\delta}(\overline C^a\chi^a +\overline D^a \overline{\chi}^a)\biggr],
  & (24) \cr
  }$$
  where $\chi^a$, $\overline{\chi}^a$ are admissible gauge fixing conditions
associated to the first class constrainst.
  We define
  $$
  \eqalignno{
  \hat{\delta}\overline C^a=-B^a\qquad &,\qquad \hat{\delta}B^a=0, & (25,a) \cr
  \hat{\delta}\overline D^a=-E^a\qquad &,\qquad \hat{\delta}E^a=0, & (25,b) \cr
  }$$
  and in order to have $\hat{\delta}S_{eff}=0$ the Lagrange multipliers
  transforms as
  $$
  \eqalignno{
  \hat{\delta} \lambda_0^a &= \wt{\cal D}_0 C^a +gf^{abc}\lambda_0^a D^c,
  & (26,a) \cr
  \hat{\delta} \omega_0^a &= {\cal D}_0 D^a. & (26,b) \cr
  }$$
  Using (22), (15,b) and (26) we see that
  $$
  \eqalignno{
  \hat{\delta}a_{\mu}^a &= {\cal D}_{\mu} D^a, & (27,b) \cr
 \hat{\delta}\omega_{\mu}^a &= \wt {\cal D}_{\mu} D^a, & (27,b) \cr
 \hat{\delta}\lambda_{\mu}^a &= \wt {\cal D}_{\mu} C^a +
 gf^{abc}\lambda_{\mu}^b D^c. & (27,c) \cr
 }$$
 We choose, now, the covariant gauge
 $$
 \eqalignno{
 \chi^a &=\mu \wt {\cal D}^{\mu} \lambda_{\mu}^a, & (28,a) \cr
 \overline {\chi}^a &= \mu \partial ^{\mu} \omega_{\mu}^a, & (28,b) \cr
 }$$
 where $\mu$ is introduced for dimensional reasons. The condition
 $\overline {\chi}^a$ is equivalent , on shell, to impose the Lorentz
 condition on $a_{\mu}$.

 The functional integral is [14]
 $$
 Z= \int {\cal D}\Phi {\left [ det \bigl\{ \psi_i^a, \psi_j^b \bigr\}
 \right ]}^{1 \over 2}\delta \bigl( \psi_i^a \bigr) \rho \quad
 exp \bigl[ S_{eff}\bigr], \eqno (29)
 $$
 where ${\cal D}\Phi \equiv {\cal D}a_i^a {\cal D}\omega_{\mu}^a
 {\cal D}\lambda_{\mu}^a {\cal D}\Pi_{i}^a {\cal D}C^a {\cal D}\overline C^a
 {\cal D}D^a {\cal D}\overline D^a {\cal D}B^a {\cal D}E^a $ is the Liouville
measure and $\rho$ is the jacobian of the functional change the variable
(15,b). The momenta
 $\Pi_i^a$ can be readily integrate out, and $a_0$ can be brought back to make
 the action covariant. We arrive finally to
 $$
 Z=\int {\cal D}a_{\mu}^a {\cal D}\omega_{\mu}^a {\cal D}\lambda_{\mu}^a
 {\cal D}C^a {\cal D}\overline C^a {\cal D}D^a {\cal D}\overline D^a
 {\cal D}B^a {\cal D}E^a \rho \quad exp \bigl[ S_{eff}\bigr],
 $$
 with $\rho=det \left (\mu^2 \delta^{ab} \delta(x) \right )$ and
 $$
 \eqalignno{
S_{eff}=\int d^3x {{\mu} \over {g^2}}tr \Biggl[ &-\mu(a_{\mu} -\omega_{\mu})
 (a^{\mu} -\omega^{\mu}) +\varepsilon^{\mu \nu \lambda}
\bigl
(\partial_{\mu} a_{\nu} a_{\lambda} +{2 \over 3}a_{\mu} a_{\nu} a_{\lambda}
\bigr)- \cr
&-{{\lambda_{\mu}} \over 2}\varepsilon^{\mu \nu \lambda}G_{\nu \lambda} -
2B \partial^{\mu}\omega_{\mu}-2E \wt {\cal D}^{\mu}\lambda_{\mu} - \cr
&-2 \overline C \wt {\cal D}^{\mu} {\cal D}_{\mu}C
-2 \overline D \partial^{\mu} \wt {\cal D}_{\mu} D-
2 \overline C \bigl[ {\cal D}^{\mu}\lambda_{\mu} ,D \bigr] \Biggr].
& (30) \cr
}$$
The effective action (30) is Lorentz covariant and $BRST$ invariant by
construction.

We have introduced a new Lorentz covariant non-abelian gauge theory describing
a spin
one massive physical mode. It is the non-abelian generalization of the
``self-dual'' abelian
theory presented in [6,7] . The dynamical germen of the evolving system  is
provided by a first
order  partial differential operator acting on the physical connection of a
principal bundle.
We have also constructed the Lorentz covariant, $BRST$ invariant  effective
action associated
to the proposed action.

It is well known the relevance of the self dual four dimencional instanton
solutions to the
 quantization of Yang-Mills Theory as well as its contribution to the
construcction of the
Topological Field Theory whose observables give rise to the Donalson invariants
of four manifolds
 directly related to the new Seiberg-Witten invariants. It would be interesting
to analyse the
role of the $2+1$ ``self-dual'' solutions from both point of view in the
context of three
dimensional field theory.

\noindent{\bf Acknowledgements}

We would like to thank Jorge Stephany for extremely helpful dicussions.

\vfill \eject
\noindent{\bf References}
\vskip 0.3cm
\item {1.}E.Witten,Comm.Math.Phys.{\bf 117}
(1988)353.
\item {2.}N.Seiberg and E.Witten,Nuc.Phys.{\bf B426}(1994)19.
\item {3.}S.Deser,R.Jackiw and S.Tempelton,Phys.Rev.Lett.{\bf 48}(1982)975;

Ann.Phys.{\bf 140}(1982)372; (E){\bf 185}(1988)406.
\item {4.}G.W.Semenoff,Phys.Rev.Lett.{\bf 48}(1988)517;

R.Mackenzie and F.Wilczek,Int. J.Mod.Phys.{\bf A3}(1988)2827;

S.Deser,Phys.Rev.Lett.{\bf 60}(1990)611.
\item {5.}A.P.Balachandran {\it et al},Int.J.Mod.Phys.{\bf A9}(1994)1569.
\item {6.}P.K.Townsend,K.Pilch and P.van Nieuwenhuizen,Phys.Lett.{\bf B136}
(1984)38;

C. R.Hagen,Ann.Phys.{\bf 157}(1984)371.
\item {7.}S.Deser and R.Jackiw,Phys.Lett.{\bf B139}(1984)371.
\item {8.}R.Jackiw and V.P.Nair,Phys.Rev.{\bf D43}(1991)1933.
\item {9.}R.Gianvittorio,A.Restuccia and J.Stephany,Mod.Phys.Lett.{\bf A6}
(1991)2121,

P.J.Arias and J.Stephany,J.Math.Phys.(1995).
\item {10.}P.J.Arias and A.Restuccia,Phys.Lett.{\bf B347}(1995)241.
\item {11.}P.A.M.Dirac,{\it ``Lectures on Quantum Mechanics''},
Belfer Graduate School of Science Monograph Series N.2
(Academic Press, 1967).
\item {12.}P.Senjanovic,Ann.Phys.{\bf 100}(1976)227.
\item {13.}I. Batalin and E. Fradkin, Phys. Lett. B122(1983) 157; Phys. Lett.
B128 (1983) 307; Ann. Inst. Henri Poincar\'e 49 (1988) 215.
 \item {14.}M.Caicedo and A.Restuccia,Class.Quan.Grav.{\bf 10}(1993) 833; Phys.
Lett. B307 (1993) 77.

\end